# Material platforms for integrated quantum photonics

S. BOGDANOV, M. Y. SHALAGINOV, A. BOLTASSEVA, AND V. M. SHALAEV*

*School of Electrical & Computer Engineering, Birck Nanotechnology Center, and Purdue Quantum Center, Purdue University, West Lafayette, IN 47907, USA*
*\* shalaev@purdue.edu*

**Abstract:** On-chip integration of quantum optical systems could be a major factor enabling photonic quantum technologies. Unlike the case of electronics, where the essential device is a transistor and the dominant material is silicon, the toolbox of elementary devices required for both classical and quantum photonic integrated circuits is vast. Therefore, many material platforms are being examined to host the future quantum photonic computers and network nodes. We discuss the pros and cons of several platforms for realizing various elementary devices, compare the current degrees of integration achieved in each platform and review several composite platform approaches.

5208–5213 (2015).

## 1. Introduction

In the 1970s and 1980s, the quantum physics community started to consider the evolution of quantum systems as a novel way of processing information. In 1984, the first eavesdropping-safe quantum key distribution (QKD) protocol was proposed [1] giving birth to the field of quantum communication [2,3]. In the 1990s, Shor [4], Deutsch [5], Grover [6], and others showed that by harnessing the classically nonexistent resource of quantum entanglement, one can achieve tremendous increase in computation efficiency compared to any classical algorithm for specific classes of problems. To achieve this, information must be encoded into quantum states of light or matter, forming quantum bits (qubits). The main obstacle for the practical realization of quantum information processing (QIP) is the decoherence of the quantum state caused by interactions with the uncontrolled environment, resulting in errors. Quantum error correction (QEC) was proven possible for quantum information [7], allowing the realization of fault-tolerant QIP. Currently, error correction only provides tolerance for minute levels of decoherence. Nevertheless, reducing decoherence in quantum systems and limiting its consequences is a fast progressing research field.

The fundamental requirements for QIP, i.e. the fine degree of control over the information carriers and the near-total isolation of these carriers from their environment, are in deep contradiction. This contradiction is embodied in the tradeoffs between typical interaction strengths and coherence times observed in basic quantum systems. Many physical realizations of qubits are currently studied, including trapped ions and atoms, superconducting circuits, quantum dots, solid-state color centers and photons [8]. Photons occupy a special place in this spectrum: they interact very weakly with transparent optical media and not at all among themselves, which renders the information they carry robust against decoherence. Due to their fast propagation speed, the use of photons is uncontested for the realization of quantum networks [9], i.e. for transmitting quantum information and distributing quantum correlations. Additionally, photons possess many degrees of freedom, which can be chosen for the encoding of quantum information. These degrees of freedom include spatial and temporal modes, frequency, polarization and angular momentum. Photons also possess continuous quantum variables, such as the quantized field quadratures, which can be utilized in special classes of information protocols [10,11]. Hybrid approaches involving both continuous and discrete variables also exist [12]. Quantum information encoded in these degrees of freedom can be manipulated with conventional optical tools. Depending on the situation, information can be conveniently transferred from one degree of freedom to another. In contrast to the manipulation of single photons, deterministic two- and three-qubit quantum gates require strong nonlinear interaction between the photons and therefore are difficult to realize. Fortunately, in 2001, Knill et al. suggested the concept of linear optics quantum computation (LOQC), where the realization of non-trivial two-photon gates is possible using linear optical elements, additional single photons and their detection [13]. Such gates only perform the desired operation in a fraction of performed attempts. This probabilistic nature makes them unsuitable for circuit-model quantum information processing [14] because of enormous overheads required to boost the success probability to the levels required for circuit-model QEC. Yet, even with moderate success probability, LOQC gates can be used to generate large multi-dimensional cluster states for subsequent measurement-based quantum computation [15,16]. In this approach, topological QEC codes are available [17], offering very advantageous fault tolerance thresholds. While fault tolerant QIP may revolutionize computation and simulation of quantum systems, elements of QIP are also required in quantum repeaters, - systems proposed to extend the range of quantum communication, which is nowadays fundamentally limited by fiber losses.

Even simple practical instances of photonic QIP e.g. those envisioned in quantum repeaters, require careful interfacing of many optical elements [18,19]. The first experimental demonstrations of quantum photonic functionalities involved tabletop optical components, which are bulky, difficult to align, stabilize and control. Nowadays, photonic QIP can leverage on the significant developments in the area of classical photonic integrated circuits (PICs) that have been realized on several material platforms. Similarly to their classical counterparts, quantum integrated photonic circuits (QPICs) promise many key advantages, including scalable, easily reconfigurable architectures, small system footprint, enhanced light-matter interaction, high stability of optical elements and the interfacing with CMOS electronics, which can perform control of optical circuits and auxiliary classical computation. Many of the stringent criteria for the production of commercial classical PICs also apply for the quantum photonic technology. However, the need to process quantum information imposes further requirements and restrictions, such as the on-demand generation of high-purity quantum optical states of light and the ultralow tolerance for losses, spectral mismatches, instabilities and detector imperfections imposed by quantum fault tolerance thresholds.

Quantum photonics has numerous applications ranging from quantum measurements to quantum communication and computation [20], but for many of them, quantum information processing and error correction are highly desirable. Therefore, the future QPICs will likely contain similar types of elementary devices, differing only by the number of required elements and device performance thresholds dictated by each specific task. The most important photonic devices for linear quantum information protocols include pump sources, non-classical light sources, filters, waveguides, directional couplers, switches, quantum memories, detectors and fiber couplers. In addition, deterministic QIP protocols require other types of devices, featuring single-photon level nonlinearities [21–23]. In QPICs, such strong nonlinearity could be enabled by solid-state quantum emitters (for a recent review see Ref [24]).

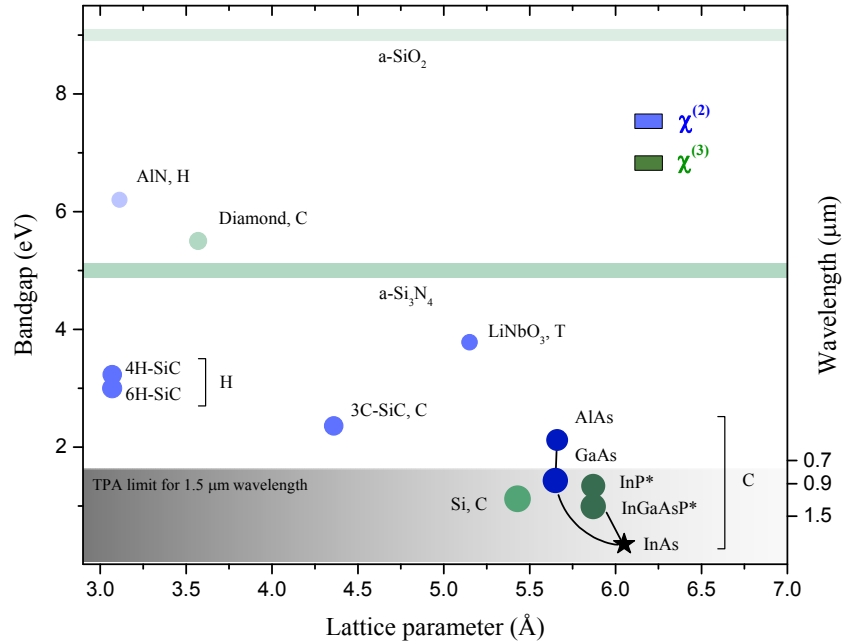

Figure 1. Properties of optical materials used in QPICs. Crystalline materials are represented by circles, while amorphous materials, ($SiO_2$ and $Si_3N_4$) are shown with stripes, since they do not possess a lattice constant. The circle diameters (and stripe thicknesses) are proportional to the refractive indices. Blue and green colors indicate second- and third-order nonlinearities respectively. The color shades qualitatively describe the strengths of nonlinear optical responses. Darker shades correspond to higher values of nonlinear coefficients. InAs is used in quantum dots and is indicated with a star. Crystal structure abbreviations: C – cubic, H – hexagonal, T – trigonal.

Because many basic devices with vastly different requirements are needed for quantum photonic chips, the realization of QPICs on various material platforms has been progressing in parallel. The objective of this review is to discuss the strengths and weaknesses of each platform and to point the reader to a number of platform- and device-specific reviews. In order to illustrate the range of opportunities offered by different materials, we will put particular emphasis on silicon-based platforms, III-V semiconductors, and diamond. In addition, we will briefly discuss other platforms, including lithium niobate, silicon carbide, silicate glasses, and plasmonic materials. In Figure 1, we present some of the most relevant optical and structural properties of these materials. We refer to platforms, made of naturally compatible materials (e.g. Si/SiO$_2$ or GaAs/AlGaAs/InGaAs) as elementary, as opposed to composite platforms, which are produced with several materials that are usually challenging to combine (e.g. Si/GaAs). We will qualitatively compare the progress achieved on elementary platforms in terms of integrating different devices and finally review the composite material approaches to achieve full on-chip functionality.

---

* InP/InGaAsP is a $\chi^{(2)}$ material system, but due to insufficient bandgap, the second-order nonlinearity is not utilized to produce correlated photons pairs at telecom wavelengths. On the other hand, second-order effects in InP/InGaAsP can be used for electro-optic modulation.

## 2. Si-based platforms

Silicon-based platforms include silica-on-silicon, silicon-on-insulator (SOI), and silicon nitride-on-silica. The first demonstrations of QPICs were made with silica-on-silicon waveguides [25]. Due to the large mode size, silica waveguides are easily interfaced with free-space optics and fibers, facilitating the first demonstrations of integrated circuits for LOQC-type quantum gates [26]. The silica-on-silicon platform is still in use today, especially for quantum communication experiments [27,28], but for larger-scale QIP purposes, it has been overshadowed by the emergence of SOI QPICs [29]. Silicon has a much higher refractive index than silica, leading to 1000 times smaller waveguide bend radius compared to that in silica waveguides. Silicon's indirect bandgap of 1.12 eV and low intrinsic carrier concentration makes it transparent to photons at the telecommunication wavelength (1.55 µm). It also features strong $\chi^{(3)}$ nonlinearity, which in combination with tighter mode confinement enables compact single-photon source designs. The SOI platform offers two major advantages over various other competing platforms: natural compatibility with the CMOS industry, and extremely well developed fabrication techniques developed for silicon electronics and photonics [30].

Single-photon production on silicon chips is based on a spontaneous four-wave mixing (SFWM) process with quantum information primarily being encoded into the photon path [31,32]. Since the individual statistics for both the signal and idler photon numbers are Poissonian, single-photon generation requires the suppression of the probabilities for obtaining zero photons and more than one photon per pulse. The vacuum component is suppressed by using one of the two photons as a herald for the spontaneous parametric generation event. The probabilities for two or more photons are reduced by attenuating the pump far below the level required to produce one pair on average. This pump attenuation makes the generation process highly probabilistic, (i.e. only one in many pump pulses will produce a photon) and strongly reduces the applicability of nonlinear single-photon sources for large-scale quantum information processing. The SWFM process can also be used to generate entangled photon pairs for quantum communication [33]. Standard telecom lasers can be used for pumping, but their on-chip integration will likely require heteroepitaxy or hybridization with another material [34]. The relatively narrow SFWM generation bandwidth in silicon complicates the pump filtering and requires a large number of highly resonant elements to achieve the desired > 120 dB pump extinction without affecting the signal and idler. By using coupled resonator optical waveguides distributed Bragg filters, currently resonant attenuations of 50 dB and more can be obtained on chip [35,36]. After generation, photons are routed using silicon wire waveguides with sub-micron mode size, promising very large-scale integration [37]. Such waveguides, in combination with modulators can realize delay lines with small programmable fixed delay ranging from tens of ps to a few ns [38–40].

Telecom range single photons cannot be detected using the well-established silicon single-photon avalanche diodes. The emergence of superconducting nanowire single-photon detectors (SNSPDs) provides an integrated high performance solution at the cost of cryogenic operation [41]. NbN and NbTiN have been so far the leading superconducting materials for on-chip waveguide integration of SNSPDs. Such detectors can exhibit a jitter in the range of tens of ps, an almost 90% quantum efficiency, and extremely low dark count rates on the order of

$10^{-4}$ Hz [42]. Potentially, superconducting detectors may incorporate photon number resolution capability either through multiplexing [43] or operation at sub-K temperatures [44]. For a recent review of waveguide-integrated SNSPDs, see [45].

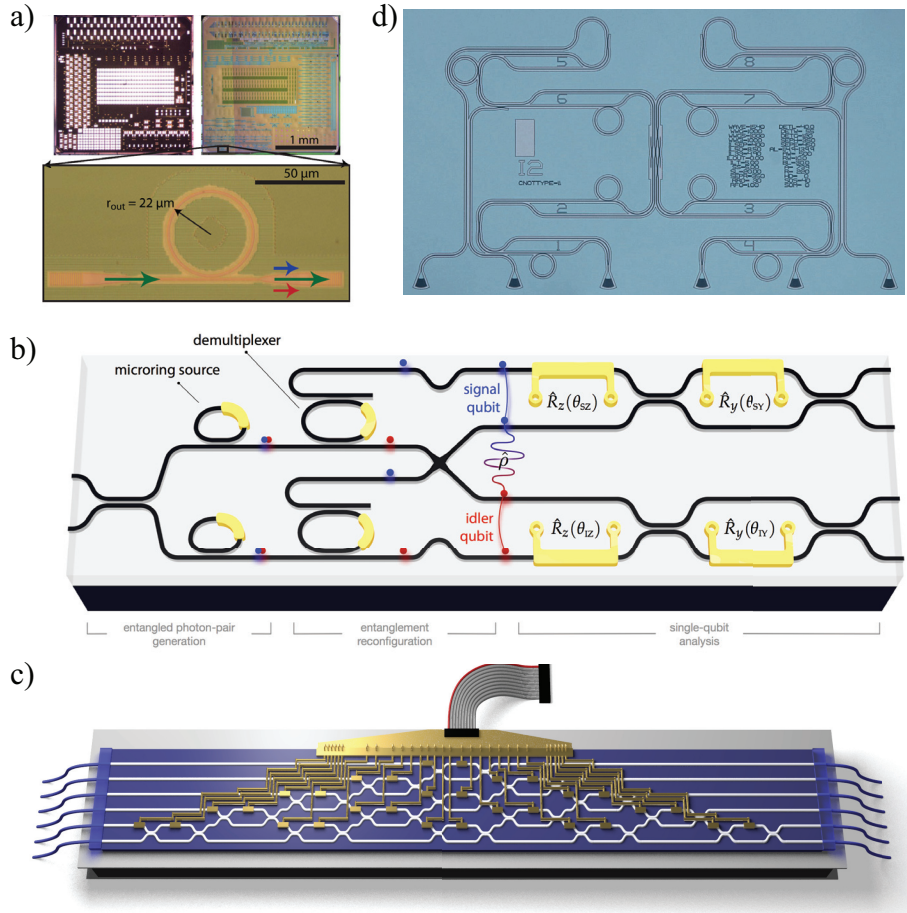

Figure 2. (a) Integration of a Si ring resonator on a 45 nm CMOS SOI chip [46]. (b) On-chip generation and analysis of a two-photon entangled state [32]. Pump laser and detectors are external to the SOI chip. (c) A fully reconfigurable circuit made with silica waveguides capable of performing all linear optical operations possible for its size [47]. (d) A micrograph showing a linear optical circuit comprising a CNOT gate and integrated elements for its characterization realized using silicon nitride-on-silica rib waveguides [48]. Figures reproduced with permission: (a) [46] from © 2015 OSA, (b) [32] from © 2015 NPG, (c) [47] from © 2015 AAAS, (d) [48] from © 2016 AIP.

Powered by the well-developed fabrication processes, the silicon-based platforms have demonstrated an almost full range of required components for the applications of quantum photonics. The integration of quantum photonic circuitry with electronics has already started using fully CMOS-compatible processes [46] (see Figure 2a). The main technological challenge now consists in scaling the number of on-chip components and achieving advanced quantum functionalities [49]. On-chip generation, entanglement and characterization of quantum states have been demonstrated for both discrete [32] (see Figure 2b) and continuous [27] variable encodings. Boson sampling circuits boast ever-increasing complexity with 15 [47] (see Figure 2c) and recently, 56 [50] fully reconfigurable interferometers on the same chip. Quantum interference-assisted simulation of simple quantum-mechanical problems has been achieved with on-chip circuitry [51].

Despite impressive demonstrations and potential, the Si-based platforms face fundamental challenges. The absence of an on-demand single-photon source on the silicon-based platforms is a serious drawback. Photonic LOQC-based QIP requires simultaneously generating many single indistinguishable photon inputs. The success of this operation falls off exponentially with the number of photons. Currently, the low probability of the single-photon generation in heralded nonlinear sources limits the number of simultaneously available photons in a quantum photonic circuit to ten [52,53]. Multiplexing many probabilistic sources could result in one deterministic source [31,54,55] but further research is needed to reduce the resource requirements for scalable integration of such sources.

Additionally, silicon is plagued by two-photon absorption (TPA) at telecom wavelengths, causing increased losses in single-photon sources. TPA also poses a problem for fast optical switching utilizing Kerr effect. The QPICs need to be quickly reconfigurable in order to allow feed-forwarding required by the measurement-based quantum computing protocols. Because of free-carrier induced losses, thermo-optic effect is generally used for modulation in QPICs rather than electrical injection and Kerr effect. However, thermo-optic modulators are too slow to allow feed-forward operations on photons stored in integrated delay lines. On the other hand, $Si_3N_4$ exhibits no TPA and, although its third-order and linear indexes are smaller than those in silicon, $Si_3N_4$ nonlinear devices can operate at higher laser powers, without sustaining increased losses [56]. For this reason, $Si_3N_4/SiO_2$ is examined as a viable candidate for QPICs. Recently, frequency conversion of single-photon level signals was achieved in $Si_3N_4$ ring resonators [57]. Small cross-section silicon-rich nitride waveguides have been introduced for switching at rates of a few GHz [58,59], providing a potentially SOI-compatible solution [60]. Using $Si_3N_4$ waveguides, integrated components for photonic quantum information have been demonstrated [48] (see Figure 2d).

## 3. III-V semiconductors

III-V based platforms, such as GaAs and InP, offer fundamentally new capabilities compared to the Si-based platforms. Well-developed semiconductor laser technology allows on-chip integration of highly tunable pump sources with electrical injection [61]. Three-dimensional confinement of both electrons and holes in heterostructures gives rise to quantum dots (QDs) which potentially provide on-demand single-photon generation across the NIR range with near-unity internal efficiency. The on-demand emission of QDs constitutes a significant advantage over heralded single-photon sources realized with nonlinear parametric processes, albeit photon indistinguishability is currently available only at cryogenic temperatures. Efficient photon collection represents a major technological challenge associated with quantum emitters. Integrated collection techniques with over 98% mode coupling efficiency exist for QDs [62], but most of them rely on complex photonic crystal structures. InGaAs/GaAs QDs demonstrate single-photon emission for wavelengths up to 1400 nm, while InAsP/InP QDs emission can be achieved across the entire telecom spectrum. Notably, when several excitons are present in the same QD, multi-photon entangled states can be generated, constituting a particularly useful resource for quantum communication. The combination of a diode junction and an embedded quantum dot has led to the demonstration of an electrically driven entangled-photon source [63] of high enough quality to perform quantum teleportation [64] (for a recent review of electrically driven quantum sources see Ref. [65]). Many III-V materials due to their lack of inversion symmetry, exhibit the electro-optic effect, which allows fast on-chip switching [66]. In addition, form birefringence can be achieved in heterostructures [67], making it possible to phase match second-order processes in III-V semiconductors for correlated photon pair generation [68].

Currently, GaAs is the most developed III-V photonic platform for quantum applications in terms of single quantum circuit elements [69]. GaAs has a high refractive index, allowing high-density integration, strong light confinement in GaAs/AlGaAs waveguides and fast electro-optic switching (see Figure 3a). Photonic crystal structures such as cavities and waveguides strongly enhance the emission speed, efficiency, purity and indistinguishability of QD-based sources as well as lead to single-photon level nonlinearities potentially useful for the realization of deterministic photon-photon interaction [70]. On-chip integration of superconducting detectors [71] has recently made it possible to generate, channel and detect single photons on the same chip [72] (see Figure 3b).

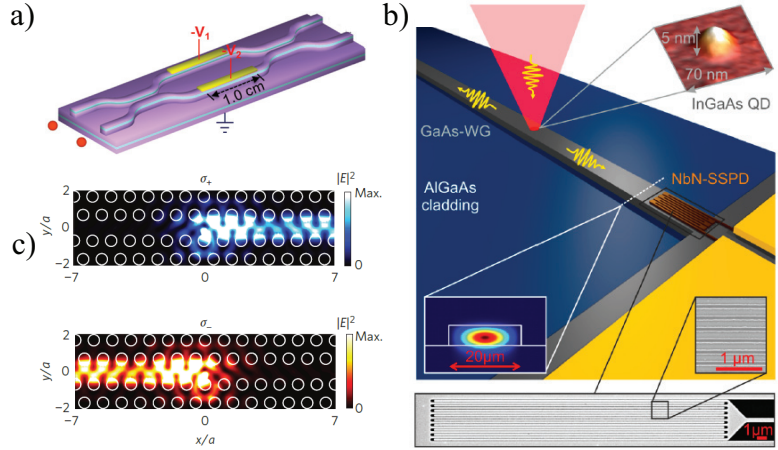

Figure 3. (a) Passive and active photonic elements realized using integrated GaAs waveguides [66]. (b) A ridge waveguide structure, hosting a layer of GaAs/InGaAs QDs and an NbN-SSPD [72]. The fluorescence from a QD is collected by the waveguide and detected on chip. The simulated waveguide mode profile, an SEM image of the SNSPD and an AFM image of a QD are shown in insets. (c) Chiral photonic crystal waveguide realizing an efficient spin-photon interface with a quantum dot for the realization of deterministic two-photon gates [73]. Figures reproduced with permission: (a) [66] from © 2014 Elsevier, (b) [72] from © 2015 ACS, (c) [73] from © 2015 NPG.

At the same time, the III-V platform faces significant challenges for producing large-scale QPICs. While deterministic on-chip QDs placement has so far produced promising results [69], achieving photon indistinguishability between different QDs still requires complicated tuning [74] or quantum frequency conversion [75] which is an active area of research by itself [76]. Finally, larger linear losses in III-V based waveguides compared to Si-based waveguides limit the achievable circuit complexities. Future research trends in III-V QPICs may include improving photon collection and detector fabrication techniques as well as managing propagation and coupling losses. Additionally, in order to apply QDs for deterministic QIP using waveguide nonlinearities [73] (see Figure 3c), as well as harnessing their spin-photon interface [77] e.g. for quantum information storage [78] significant suppression of the decoherence mechanisms for spin and fluorescence in QDs is needed. Additionally, integration with on-chip electronics will be very beneficial for unlocking the full power of fault-tolerant quantum information processing.

## 4. Bulk diamond and diamond-on-insulator

Diamond has long been known as a material with exceptional mechanical, thermal and optical properties but difficult to synthesize and process. Thanks to the recent development of a plethora of fabrication techniques [79], diamond has emerged as a competitive platform for realizing highly functional QPICs [80]. The unique advantages of diamond in this area stem from its wide bandgap, high Debye temperature, high isotopic purity and low free electron concentration. This material brings to the table hundreds of mid-gap defects [81], some of which constitute outstanding quantum systems. Mainly, two types of diamond platforms are in use. Bulk single-crystal diamond constitutes the first platform, and is preferred in situations where color centers with superior properties are desired. However, bulk diamond photonic circuits are difficult to process because of the need for freestanding optical structures. Another approach consists in growing polycrystalline diamond films [82,83] on or hybridizing thin crystalline diamond membranes [84,85] onto insulator substrates. The diamond-on-insulator or DOI platform constitutes a relatively fabrication-friendly alternative to bulk diamond substrates.

Tremendous progress has been achieved in the theoretical investigation, synthesis, identification, characterization and quantum state manipulation of color centers. Nitrogen-vacancy (NV) centers [86] feature bright and stable fluorescence and possess unrivaled electron spin coherence times even at room temperature, which finds applications in nanoscale sensing [87]. Optical single-shot spin readout and coherent manipulation is possible at cryogenic temperatures [88]. Single nuclear spins in diamond possess a coherence time on the order of seconds. When located in the direct vicinity of an NV center, they can be harnessed through hyperfine interaction with the NV electronic spin. Utilizing nuclear spins has already been shown to result in storage of quantum information [89], quantum error correction [90,91] (see Figure 4a), and coherent transfer of quantum information from a photon to a nuclear spin [92]. These functionalities make the NV center a fully functional quantum memory and a register for fault-tolerant QIP.

Silicon-vacancy (SiV) centers, due to their orbital symmetry possess lifetime-limited optical transitions with no spectral diffusion [93], very small inhomogenous linewidths [94] and near-unity Debye-Waller factor [95]. These color centers can be potentially used for scalable on-demand generation of indistinguishable single photons without need for frequency tuning [96]. The finite intrinsic quantum efficiency of the SiV (up to 60%) can be improved e.g. through moderate Purcell effect [97]. Although the SiV electronic spin has a much shorter coherence time than that of the NV center, and requires substantial magnetic field for optical addressing [98], its additional metastable orbital states can be optically controlled in ultrafast manner [99] to achieve quantum optical switching and on-chip SiV-SiV entanglement [100]. More recently, Cr- [101], Ni- [102], Ge- [103] and Xe-related [104] centers have been attracting attention for their promising optical properties. In a few years, they are expected to find their respective roles in the diamond color center family.

Recently, spin-photon entanglement in NV centers [105] has been leveraged to demonstrate advanced quantum network capabilities, including short distance quantum teleportation [106] and entanglement of remote spins [107]. Diamond waveguide structures (see Figure 4b) exhibit total internal reflection in wide range of angles and naturally provide an integrated highly efficient interface for color center manipulation and fluorescence collection. In combination with diamond waveguides and photonic crystal cavities, the color centers in diamond could open the avenue for the integrated deterministic generation and manipulation of quantum information [108]. This approach is to be contrasted with the inherently probabilistic operation of linear QPICs and could significantly reduce the infrastructural overheads for achieving large-scale fault-tolerant QIP.

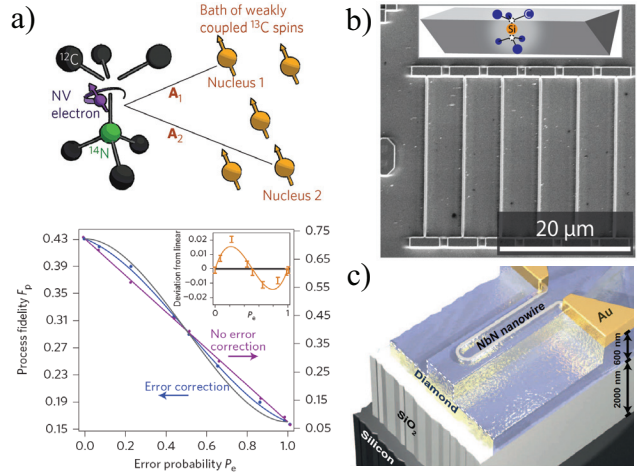

Figure 4. (a) Diamond lattice nuclear spins neighboring a single NV center (top) allow the realization of quantum error correction (bottom) [90]. (b) Free-standing nanobeam diamond waveguides containing SiV centers with narrow inhomogeneous spectral distribution for scalable single-photon generation [94]. (c) NbN SNSPD integrated onto a diamond-on-insulator rib waveguide featuring an on-chip detection efficiency of 66% [109]. Figures reproduced with permission: (a) [90] from © 2014 NPG, (b) [94] from © 2016 APS, (c) [109] from © 2015 CIOMP.

Diamond patterning and deterministic color center creation techniques are rapidly progressing. Free-standing photonic crystal cavities have been fabricated in diamond membranes grown on sacrificial substrate [110]. NbN-based SNSPDs have been fabricated on polycrystalline diamond surface grown via chemical vapor deposition (CVD) on a silica-on-silicon wafer [109,111] (see Figure 4c). Electro-optomechanical The recently demonstrated femtosecond laser based bulk diamond writing technique [112] may hold potential for flexible waveguide definition. Contrary to III-V QDs, the most promising color centers in diamond emit at wavelengths below 1 μm, which hurdles the prospects of long distance communication. More importantly, no large-area single-crystal thin film or membrane has so far been achieved. Better understanding of the ion implantation process is needed to improve color center yield and obtain better control of dipole orientations. Integration of auxiliary control electronics seems challenging on both bulk diamond and DOI platforms. Nevertheless, the unique and diverse functionalities of color centers motivates the continuing research into diamond growth, processing and implantation techniques [80], which will make diamond a strong contender for integrated quantum photonics in the years to come.

## 5. Other platforms

The materials used for QPICs are very diverse. The above mentioned silicon-, III-V semiconductors- and diamond-based platforms aims to paint a representative picture of the various possibilities offered by these materials. However, basic integrated quantum photonic functionalities have been demonstrated in other material systems.

Lithium niobate (LN) is a highly versatile optical material, used since the very first days of integrated quantum optics [113]. Its valuable properties include strong $\chi^{(2)}$ nonlinearity, birefringence, electro-optic effect, ferroelectricity, high transparency and chemical stability. Large-scale wafers are produced commercially and fabrication techniques are well developed [114]. Waveguides can be formed by proton exchange or titanium diffusion techniques [115,116] and periodically poled to greatly increase the efficiency of the spontaneous parametric generation process. High quality correlated photon pair generation and interference [117] (see Figure 5a), fast electro-optic modulation [118] have been achieved in on-chip LN waveguides. By harnessing nonlinear processes in coupled waveguide structures [119], one can generate optically reconfigurable quantum states [120,121] and achieve strong spatial pump filtering [122]. Embedding of rare-earth ions is being investigated for deterministic quantum memory applications [123]. However, LN devices are often large in size, and the experiments are mostly limited to quantum communication-related functionalities [124].

Silicon carbide (SiC) is a high bandgap group IV material, which, like diamond, offers a wide variety of color centers. In contrast to diamond, fabrication techniques for SiC structures are better developed [125,126]. These color centers can serve as bright and stable single-photon sources [127,128] and optically addressable spin qubits with coherence times approaching milliseconds [129,130]. When integrated into bulk SiC optical resonators, (see Figure 5b) color centers exhibit stronger fluorescence and spin readout signals [131]. The research on SiC color centers is in its relatively early stages. For example, so far no color center exhibiting both bright fluorescence and long spin coherence time has been identified. Still, characterization of new color centers in SiC is rapidly progressing, and the potential of this material for quantum photonics has yet to be fully appreciated.

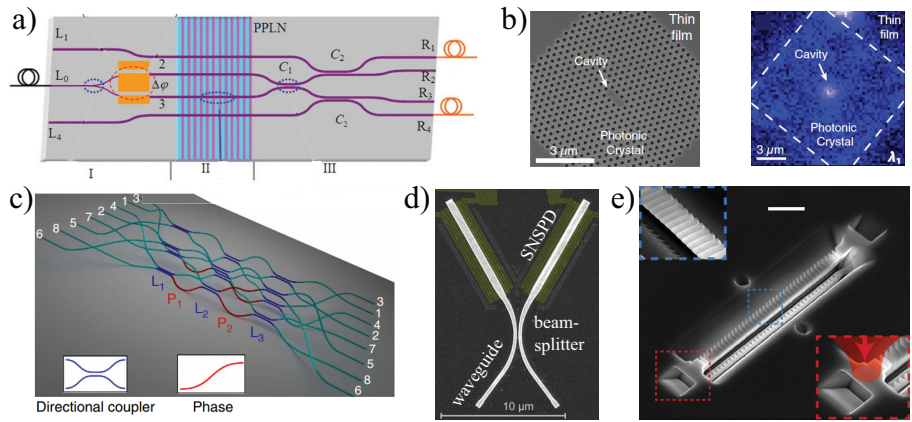

Figure 5. (a) A lithium niobate chip with nonlinear, passive and active elements, allowing generation and characterization of entangled photon pairs [117]. (b) SEM image (left) and confocal PL scan (right) of a SiC-on-Si photonic crystal cavity containing an ensemble of Ky5 spin defects [131]. (c) Schematic of a three-dimensional 8-mode interferometer written by laser beam in borosilicate glass, realizing transformations described by Fourier matrices [132]. (d) A directional coupler realized with two gold plasmonic waveguides. NbN SNSPDs are integrated on the waveguides ends for realizing an on-chip two-photon interference experiment [133]. (e) SEM scan of a nanobeam YSO resonator containing $Nd^{3+}$ ions for the realization of an optical quantum memory [134]. Figures reproduced with permission: (a) [117] from © 2014 APS, (b) [131] from © 2013 NPG, (c) [132] from © 2013 Wiley-VCH, (d) [133] from © 2015 NPG, (e) [134] from © 2016 NPG.

Optical circuits produced by femtosecond laser direct writing in silicate glasses (see Figure 5c) offer interesting possibilities for quantum simulation and for emulating complex quantum operators [135]. This fabrication technique does not require any pre- or postprocessing of the substrates. Large-scale production of waveguides and passive devices is possible in three dimensions with writing speeds up to cm/s range. The integration of standard QPIC components such as single-photon sources and detectors as well as the control electronics is still an outstanding issue. Additionally, relatively small refractive index contrast in the waveguides prevents high density integration. Nonetheless, silicate glass circuits obtained by direct writing feature fast production, high fabrication tolerance, and great design flexibility, including the possibility to manipulate polarization-encoded photons and realize three-dimensional waveguide arrangements. Such circuits are very attractive for chip prototyping and fundamental complexity tests.

While dielectric-based optics is fundamentally diffraction-limited, higher mode confinement can be obtained by coupling photons to collective electron excitations at metal-dielectric boundaries, resulting in surface plasmon-polaritons (SPPs). In optical communications, plasmonics is considered a promising route towards on-chip high bandwidth optical signal routing infrastructure, matching the size of electronic components [136]. For quantum photonics, plasmons are interesting in particular thanks to their ability to strongly couple to even broadband quantum emitters [137,138], opening the route for deterministic room-temperature QIP. Despite several years of intense efforts [139], quantum plasmonics remains a relatively unexplored area of research. Preservation of quantum coherence in plasmonic circuits has been observed in several on-chip experiments [140–143]. On-chip two-plasmon interference [144] and superconducting plasmon detector integration [133] have also been demonstrated (see Figure 5d). Plasmonics is plagued by losses [145], which seriously limit most quantum photonic applications. However, new plasmonic materials such as epitaxial silver [146], epitaxial gold [147] and transparent conductive oxides (TCOs) [148] offer lower loss than polycrystalline noble metals, and, in the case of TCOs, high degree of tunability [149], high third-order nonlinearity [150,151], and CMOS-compatibility. The main question is whether the loss can be overcome e.g. using gain media [152], but the gradual emergence and development of alternative plasmonic platforms could unlock the door to nanoscale quantum optical devices with unique properties.

The use of rare-earth ions implanted into various substrates has been receiving increasing attention for implementing quantum memories. Notably, the use of $Y_2SiO_5$ substrate leads to exceptionally high rare earth ion coherence times [153], recently demonstrated to exceed six hours [154]. Rare-earth doped $Y_2SiO_5$ have been used to demonstrate efficient quantum information storage [155] and operation at telecom wavelength [156]. Integration of such quantum memories onto a chip has been prototyped using the direct writing technique [157] and free-standing nanobeam waveguides [134] (see Figure 5e), resulting in frequency-encoded qubit storage with fidelity approaching 99% [158]. For a recent review of quantum memories, see Ref. [38].

## 6. Device integration

While the development of single quantum photonic devices is very intense across many material platforms, the efficient integration of these devices on the same chip is required to unlock the power of integrated quantum photonics. Therefore, we propose a qualitative comparison of different elementary platforms based on the degree of integration reported in literature. Table 1 shows examples of integrated chips where we have attempted to capture at the same time a spectrum of different devices and the highest degree of integration achieved to date. Each platform is represented by a different color and each line in the table corresponds to the devices demonstrated within a single paper. The filled cells mark the availability of the corresponding device and the darker shaded cells indicate the devices that are particularly suitable for quantum applications.

Such comparison is necessarily subjective and does not include all the platforms mentioned in this review. Si and $LiNbO_3$ display a particularly high degree of integration, given their long history of development as photonic materials. Based on these platforms, on-chip systems consisting of four or more different categories of devices have been demonstrated. We remark, however, that no system up to date demonstrates on-chip integration of all components. Moreover, amongst the elementary platforms, perhaps, only the III-V materials currently offer the entire range of required on-chip devices ranging from a pump laser to single-photon detectors. They therefore appear the most likely to achieve full integration within the next few years.

**Table 1. State-of-the-art integration of quantum photonic devices in on-chip circuits.** Color shades qualitatively indicate the performance of each device, with darker shaded cells corresponding to better devices.

| | Pump | SPS | CPPS | Waveguide | Passive elements | Active elements | Detector | Fiber coupler |
|---|---|---|---|---|---|---|---|---|
| SOI [159] | | | SFWM 1.5µm | strip Si/SiO$_2$ | MMI, WGX | TOPS | | Grating |
| SOI [42] | | | | strip Si/SiO$_2$ | YS, DL | | NbN wire | Grating |
| SiO$_2$ [44] | | | | Ge:SiO$_2$/SiO$_2$ laser written | | | TES | Pigtailing |
| SiO$_2$ [47] | | | | Ge:SiO$_2$/SiO$_2$ | DC | TOPS | | Pigtailing |
| SiO$_2$ [160] | | | SFWM 0.7µm | SiO$_2$ laser written | | | | |
| Si$_3$N$_4$ [161] | | | | TriPleX | DC, DL | TOPS | | pigtailing |
| Si$_3$N$_4$ [162] | | | SFWM 1.5µm | strip Si$_3$N$_4$/SiO$_2$ | | | | |
| Si$_3$N$_4$ [163] | | | | strip Si$_3$N$_4$/SiO$_2$ | DC, YS | | NbTiN wire | grating |
| InP [164] | SOA 1.5 µm | | | ridge & strip InGaAsP/InP | DC | EOPM | | |
| GaAs [72] | | QD InGaAs/GaAs 1µm | | strip GaAs/AlGaAs | | | NbN wire | |
| GaAs [64] | | | ELED InAs/GaAs 0.9µm | | | | | |
| Diamond [100] | | SiV 0.7 µm | | free-standing wire | | | | adiabatic mode transfer |
| DOI [165] | | | | ridge diamond/SiO$_2$ | YS | | NbN wire | Grating |
| LiNbO$_3$ [117] | | | SPDC PPLN 1.5µm | strip H+ exchange in LiNbO$_3$ | DC | EOPM | | fiber tips |
| LiNbO$_3$ [166] | | | | strip H+ exchange in LiNbO$_3$ | | | NbN wire | |

---

CPPS – correlated photon pair source, DC – direct coupler, DL – delay line, DOI – diamond-on-insulator, ELED – entangled light emitting diode, EOPM – electro-optic phase modulator, MMI – multimode interference coupler, MQW – multiple quantum wells, QD – quantum dot, SFWM – spontaneous four-wave mixing, SiV – silicon-vacancy center, SOA – semiconductor optical amplifier, SPDC – spontaneous parametric down-conversion, SPS – single-photon source, TOPS – thermo-optic phase shifter, WGX – waveguide crossing, YS – Y-split coupler.

## 7. Beyond single-platform approaches

Besides efforts directed at demonstrating full QPIC functionality on a single platform, various material combinations have been explored through hybrid and monolithic integration approaches. Such combinations are motivated by strategic complementarities in material properties and are sometimes a straightforward development of analogous demonstrations and technologies in classical photonics. For some applications, such as weak coherent state-based QKD, active and passive functionalities can be allocated to different chips [164]. However, for most quantum information protocols, composite material platforms may provide the only realistic fully integrated solutions. We discuss below several examples of such integration, focusing on silicon as the host substrate.

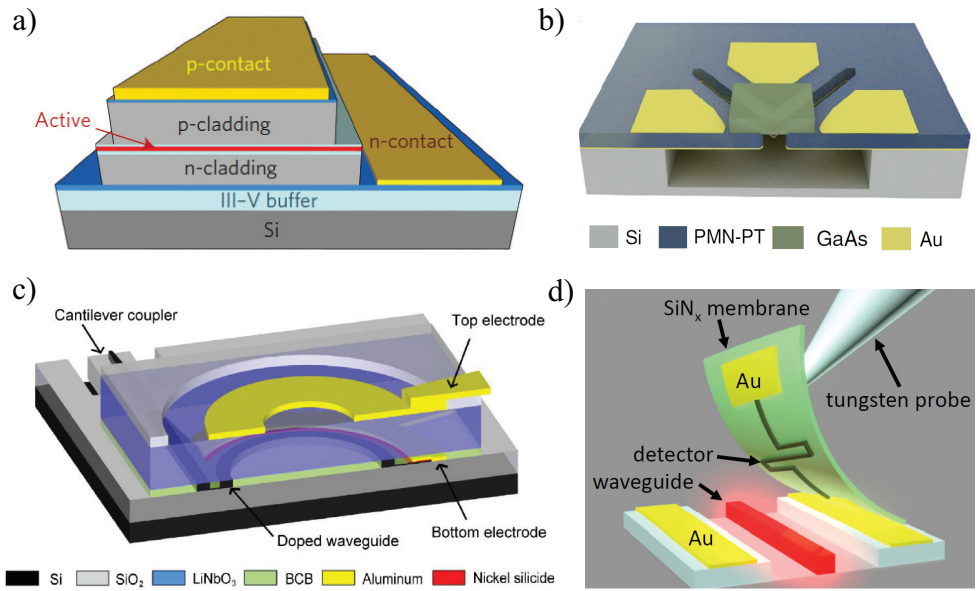

Figure 6. (a) Structure of an electrically injected InGaAs/GaAs QD 1.3 μm laser grown on silicon [167]. (b) Schematic of a wavelength tunable entangled-photon source integrated on Si, employing InAs/GaAs QDs and $[Pb(Mg_{1/3}Nb_{2/3})O_3]_{0.72}[PbTiO_3]_{0.28}$ actuators [168]. (c) Electro-optical ring modulator realized by bonding lithium niobate thin film onto a processed silicon chip [169]. (d) High-yield fabrication of SNSPD detectors by hybridizing precharacterized NbN nanowires on silicon nitride membranes with AlN and silicon waveguides [170]. Figures reproduced with permission: (a) [167] from © 2014 OSA, (b) [168] from © 2016 NPG, (c) [169] from © 2015 NPG, (d) [170] from © 2016 OSA.

Silicon is not well-suited for lasing because of its indirect bandgap. In contrast, III-V materials excel at lasing and switching but have less developed fabrication processes for both CMOS electronics and on-chip waveguides. Recently, optically [171] and electrically [167] pumped InAs/GaAs quantum dot lasers (see Figure 6a) have been demonstrated by direct growth on Si substrates working at 1.3 μm and an InP laser array has been integrated on Si [172], although for the latter, electrical pumping has not yet been achieved. Hybridization via wafer bonding is currently the most robust albeit technologically involved approach towards III-V-on-Si lasers [34]. It is not limited by alignment issues and lattice mismatch that respectively plague directly bonded devices and heteroepitaxy methods, but requires either a careful engineering of a spacer layer or atomically pure wafer surfaces. Other routes towards achieving high power electrically-injected lasers on Si platforms include bandgap engineering in Ge grown on Si and using rare-earth-doped silicon-rich materials as active medium [173].

Integration of nonclassical light sources onto Si-based platforms is an attractive alternative to multiplexed heralded sources. Heteroepitaxial growth of III-V quantum dots on Si is a promising direction for efficient generation of single photons [174]. Additionally, Murray et al. [175] demonstrated orthogonal bonding of a GaAs chip containing QDs and a silicon photonic chip with single-qubit circuitry. Probe-based manipulation has allowed integrating QDs into SiN waveguides [176] and onto MEMS tuners [168] (see Figure 6b), as well as positioning nanodiamond-based NV centers into photonic crystal structures [177] and plasmonic waveguides [178]. Diamond microwaveguides and PCCs containing single NV centers have been hybridized onto silicon chips [179,180], promising high quality hybrid quantum memories and single-photon sources (especially in the case of SiV centers). Such approaches are naturally difficult to scale, but offer deterministic integration of precharacterized emitters and devices. For a review of nanomanipulation-facilitated hybrid nanophotonic structures see [181]. Single-photon emission from intrinsic [182] and doped [183] carbon nanotubes (CNTs) has led to efforts in constructing scalable hybrid photonic circuits. Large-scale deterministic placement of CNTs by dielectrophoresis [184] was recently utilized to construct integrated QPICs featuring on-chip generation, guiding and detection of single photons [185].

Fast control of weak optical signals by electro-optic effect on Si platform without introducing loss has motivated integration of $\chi^{(2)}$ media on silicon chips. Such media range from organic electro-optic materials [186,187] and lithium niobate [169,188] (see Figure 6c) to III-V semiconductors [189] and ferroelectric perovskites [190]. Such integration potentially offers gigahertz electronic modulation bandwidth. Other nonlinear platforms utilizing silicon substrates such as AlN [191,192] and Hydex [56,193] demonstrate interesting properties for applications in quantum networks. A hybrid approach has also been utilized to increase SNSPD fabrication yield, by bonding SiN membranes with precharacterized superconducting lines to silicon and AlN waveguides [170] (see Figure 6d).

In order to realize deterministic generation and manipulation of quantum information, efficient feed-forwarding and memory functionalities, while embedding classical processing and control circuits, the use of several optical materials on the same chip seems hardly avoidable. Device hybridization is attractive to demonstrate high quality individual devices with high fabrication yield, but is a slow, costly and therefore lesser desirable procedure for industry-scale production. Composite wafers can be obtained by wafer bonding, but the resulting devices suffer from lack of uniformity and yield. Arguably, heteroepitaxy may still be the preferred route towards achieving full functionality in complex programmable QPICs. This motivates further research of growth and processing techniques which can be useful far beyond quantum science and technology.

## 8. Conclusion

The future of integrated quantum photonics relies on the exploitation of fundamental possibilities offered by various materials. Many material platforms have demonstrated remarkable potential in terms of both hosting a full range of devices required for QIP and large-scale integration. Nevertheless, major challenges remain at the single device level. Amongst the most pressing issues are the fabrication of on-demand indistinguishable single-photon sources and the integration of high performance single-photon detectors, preferably operating at or near room temperature. Integrated quantum memories with high delay-bandwidth product for telecom range photons are of particular importance for quantum communication. Because of strong requirements imposed on materials and a great variety of devices needed to implement highly functional QPICs, it looks likely that a composite material approach will be very important if not unavoidable. In this regard, monolithic approaches to integration of III-V devices on silicon look particularly promising. The growing availability of foundry services and rapid advances in classical photonics have supported the explosive progress in photonic quantum technologies. In return, the design and realization of large scale QPICs, if successful, will produce a profound impact on science and technology through advances in information and computation theory, material engineering, as well as achievement of fundamental operation limits in integrated optical and optoelectronic devices.

**Funding**

Air Force Office of Scientific Research (AFOSR) MURI (29017320-51649-D, FA9550-14-1-0389), National Science Foundation (NSF) MRSEC (DMR-1120923).

**Acknowledgements**

The authors thank Jungu J. Choi, Gregory. N. Goltsman, Mahdi Hosseini, Vadim. Kovalyuk, Patrik Rath, Shreyas. Shah and Alexander S. Solntsev for fruitful discussions, Jacques Carolan, Reinier W. Heeres, Menno Poot, Joshua W. Silverstone, Alp Sipahigil, Tim H. Taminiau, Jianwei Wang, Ping Xu, and Tian Zhong for assistance with images, and Samuel Peana for his help with manuscript preparation.